\documentclass[aps,prl,showpacs,twocolumn,superscriptaddress]{revtex4}

\usepackage{graphicx}
\usepackage{bm}
\begin{document}
\title{Interplay between magnetic anisotropy and interlayer coupling in nanosecond magnetization reversal
of spin-valve trilayers}

\author{J.~Vogel}
\affiliation{Laboratoire Louis N\'{e}el, CNRS, 25 avenue des
Martyrs, B.P. 166, 38042 Grenoble Cedex 9, France}
\author{W.~Kuch}
\altaffiliation[Present address: ]{Institut f\"{u}r
Experimentalphysik, Freie Universit\"{a}t Berlin, Arnimallee 14,
14195 Berlin, Germany.} \affiliation{Max-Planck-Institut f\"{u}r
Mikrostrukturphysik, Weinberg 2, 06120 Halle, Germany}
\author{J.~Camarero}
\affiliation{Dpto. F\'{i}sica de la Materia Condensada, Universidad
Aut\'{o}noma de Madrid, 28049 Madrid, Spain}
\author{K.~Fukumoto}
\altaffiliation[Present address: ]{Institut f\"{u}r
Experimentalphysik, Freie Universit\"{a}t Berlin, Arnimallee 14,
14195 Berlin, Germany.} \affiliation{Max-Planck-Institut f\"{u}r
Mikrostrukturphysik, Weinberg 2, 06120 Halle, Germany}
\author{Y.~Pennec}
\altaffiliation[Present address: ]{Department of Physics, University
of Alberta, Edmonton, Alberta, Canada T6G 2J1.}
\affiliation{Laboratoire Louis N\'{e}el, CNRS, 25 avenue des
Martyrs, B.P. 166, 38042 Grenoble Cedex 9, France}
\author{S.~Pizzini}
\affiliation{Laboratoire Louis N\'{e}el, CNRS, 25 avenue des
Martyrs, B.P. 166, 38042 Grenoble Cedex 9, France}
\author{M.~Bonfim}
\affiliation{Departamento de Engenharia El\'{e}trica, Universidade
do Paran\'{a}, CEP 81531-990, Curitiba, Brazil}
\author{F.~Petroff}
\affiliation{Unit\'{e} Mixte de Physique CNRS/Thales, Domaine de
Corbeville, 91404 Orsay, France, and Universit\'{e} Paris-Sud, 91405
Orsay, France}
\author{A.~Fontaine}
\affiliation{Laboratoire Louis N\'{e}el, CNRS, 25 avenue des
Martyrs, B.P. 166, 38042 Grenoble Cedex 9, France}
\author{J.~Kirschner}
\affiliation{Max-Planck-Institut f\"{u}r Mikrostrukturphysik,
Weinberg 2, 06120 Halle, Germany}

\date{December 25, 2004}

\begin{abstract}
The influence of magnetic anisotropy on nanosecond magnetization
reversal in coupled FeNi/Cu/Co trilayers was studied using a
photoelectron emission microscope combined with x-ray magnetic
circular dicroism. In quasi-isotropic samples the reversal of the
soft FeNi layer is determined by domain wall pinning that leads to
the formation of small and irregular domains. In samples with
uniaxial magnetic anisotropy, the domains are larger and the
influence of local interlayer coupling dominates the domain
structure and the reversal of the FeNi layer.
\end{abstract}


\maketitle

Magnetic trilayers in which two thin ferromagnetic films are
separated by a non-magnetic spacer layer present a variety of
effects - giant magnetoresistance, tunnel magnetoresistance, spin
torque transfer - that make them highly interesting for both
fundamental studies and applications. Recent studies of
magnetization dynamics of trilayer systems like spin valves (SV) and
magnetic tunnel junctions (MTJ)
\cite{Schum03,Bonfim01,Kaka02,Heinrich03,Pennec04} are mainly
motivated by magnetic recording and memory applications, since the
switching speed of their active magnetic element, the soft
ferromagnetic film, can ultimately limit the rate at which
information can be read or written in the devices. At nanosecond
timescales, the magnetization reversal of ferromagnetic layers is
determined by processes like nucleation and domain wall propagation
that are strongly sensitive to the magnetic anisotropy of the film.
In magnetically coupled trilayers, also the interaction between the
magnetic layers influences the reversal. Despite the fundamental
interest of these effects and the consequences for technological
applications, few studies have been published on the influence of
interlayer coupling and anisotropy on the fast magnetization
reversal of magnetically coupled trilayers
\cite{Fassb01,Hicken03,Layadi03}. In this paper, we show that the
magnetic anisotropy within the plane of the layers has a large
influence on the shape of the magnetic domains and on the domain
wall dynamics, as well as on the correlation between the domain
structures in the soft and hard magnetic layers. In quasi-isotropic
samples the application of nanosecond magnetic pulses gives rise to
small and irregular domains in the soft layer. This leads to a large
density of $360^{\circ}$ domain walls and consequently to a large
increase of the saturation field. This phenomenon can be avoided
using layers with a uniaxial magnetic anisotropy.

We have independently studied the nanosecond magnetization reversal
of the soft permalloy (Fe$_{20}$Ni$_{80}$) and of the hard Co layer
in spin-valve like FeNi/Cu/Co trilayers using time and
layer-resolved photoelectron emission microscopy (PEEM) combined
with x-ray magnetic circular dichroism (XMCD) \cite{schneider}. In
XMCD-PEEM secondary electrons emitted from the sample surface after
resonant absorption of circularly polarized x-rays are collected in
an electron microscope to obtain an image of the magnetic domain
structure. The magnetic contrast is caused by the difference in
x-ray absorption of magnetic domains having their magnetization
parallel or anti-parallel to the direction of the incoming
circularly polarized x-rays. X-ray imaging techniques have been used
to investigate magnetization dynamics mainly in permalloy structures
\cite{vogel,trx}, but the chemical selectivity that can provide
layer-resolved imaging has been little exploited
\cite{vogelMMM,KuchAPL}.

The measurements were performed at the UE52-SGM and the UE56/2-PGM2
helical undulator beamlines of the BESSY synchrotron radiation
source in Berlin, Germany. Temporal resolution was obtained using a
pump-probe scheme, like in our previous time-resolved XMCD
measurements \cite{Bonfim01}. Magnetic pulses provided by a
microcoil were synchronized with the photon bunches at a repetition
rate of 625 kHz and images were acquired with different delays
between the magnetic and photon pulses. The acquisition time per
image was 5 to 10 minutes, corresponding to an average over several
hundreds of millions of pulses. More details can be found in Ref.
\cite{vogel}. The set-up of the electrostatic photoelectron emission
microscope (Focus IS-PEEM) was described in Ref. \cite{kuch98}. The
angle of incidence of x-rays on the sample was 60${^{\circ}}$ from
the surface normal. The photon energy was tuned to the Fe-L$_3$ (Co
L$_3$) absorption edge to image the permalloy (Co) domain structure.
The electrons emitted by the Co layer had to travel through 12 nm of
other material to reach the surface, leading to a weaker contrast
for the Co images.

\begin{figure}
\includegraphics*[bb= 147 378 384 480]{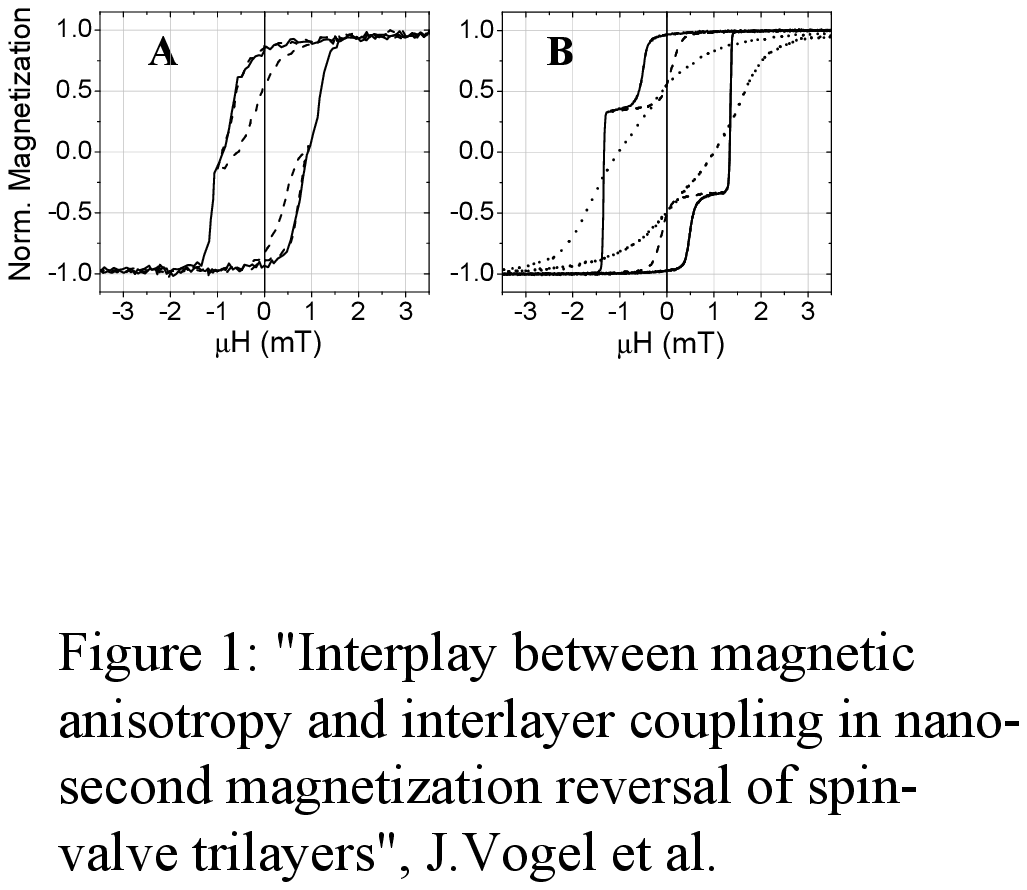}
\caption{Major and minor hysteresis loops obtained by longitudinal
Kerr effect for two Fe$_{20}$Ni$_{80}$(5nm)/Cu(4nm)/Co(\textit{x}
nm) spin-valve-like trilayers deposited on Si(001)/SiO$_2$. Sample A
(\textit{x}=5) is quasi-isotropic within the plane of the layers
while sample B (\textit{x}=8) presents uniaxial magnetic anisotropy.
Minor loops of the FeNi layer are indicated by dashed lines. For
sample B, both the in-plane easy (continuous line) and hard (dotted
line) magnetization axis loops are shown.}
\end{figure}

Two different Fe$_{20}$Ni$_{80}$/Cu/Co trilayers deposited on
SiO$_2$/Si(001) substrates by RF sputtering were studied. The
thickness of the permalloy and Cu layers, 5~nm and 4~nm
respectively, was the same for both samples. The Co thickness was
5~nm for sample A and 8~nm for sample B. The samples were capped
with an Al layer of 3~nm to protect them from oxidation.
Quasi-static hysteresis loops obtained using longitudinal
magneto-optical Kerr effect are shown in Figure~1. The magnetization
of both samples was in the plane of the layers. Hysteresis loops
taken with the magnetic field applied along several azimuthal angles
revealed no clear magnetic anisotropy within this plane for sample
A. A small in-plane magnetic field applied during the growth of
sample B led to a uniaxial magnetic anisotropy for the permalloy
layer, with the easy axis parallel to the applied field direction.
Both samples show two transitions at different fields, corresponding
to the separate reversal of the permalloy (lower coercivity) and Co
layers. The minor loops of the permalloy layer (indicated by dashed
lines in Fig.~1) are shifted with respect to zero field due to a
coupling with the Co layer of about 0.4 mT in both samples. This
magnetostatic coupling is due to correlated roughness at the two
ferromagnetic/non-magnetic interfaces \cite{neel}. The hysteresis
loops are more tilted for sample A. As we will see later, this
indicates a lower barrier for nucleation than for sample B and a
larger influence of the pinning of domain wall motion.

\begin{figure}
\includegraphics*[bb= 123 459 371 608]{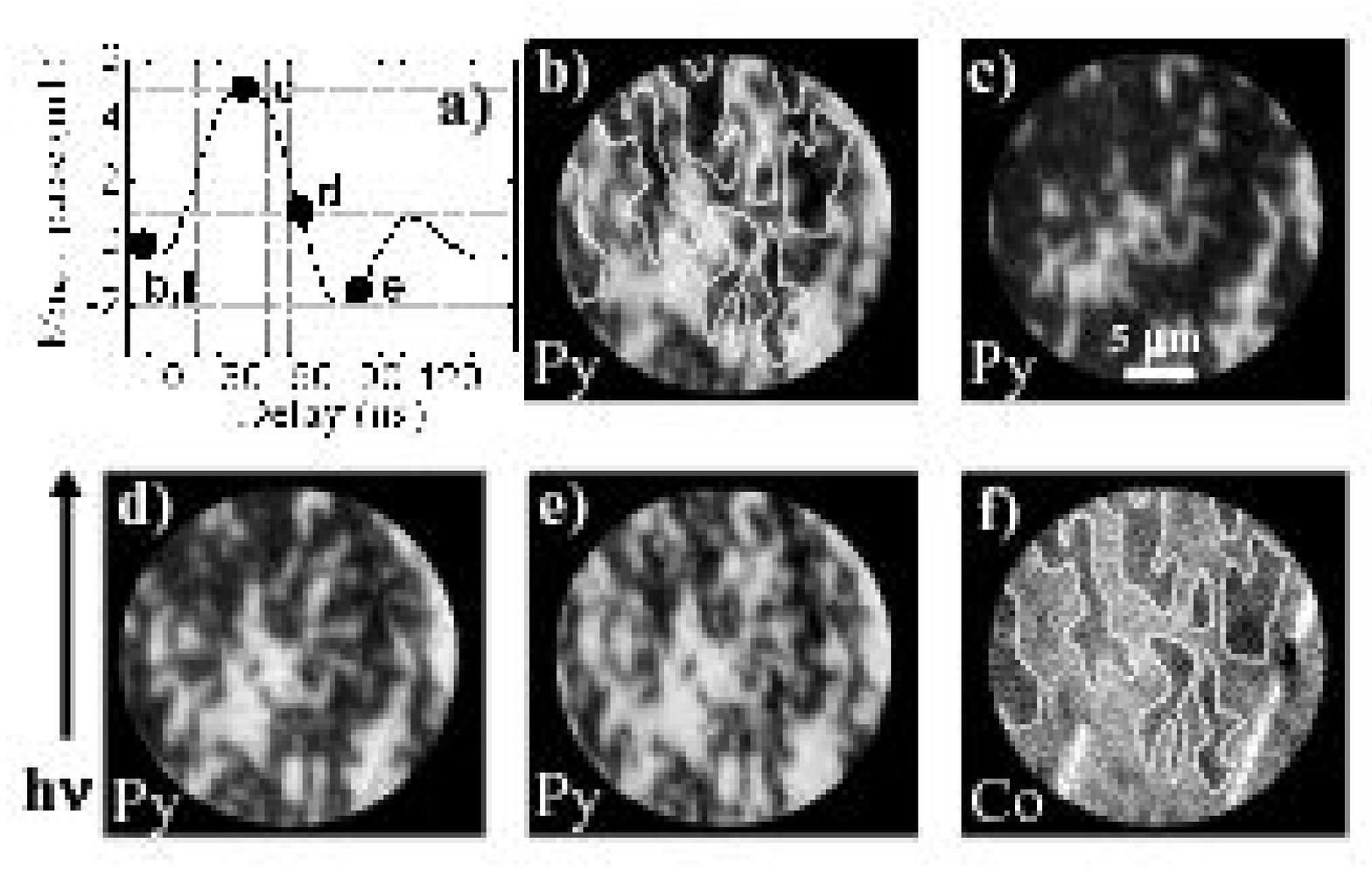}
\caption{Time and layer-resolved XMCD-PEEM domain images of the
permalloy [(b)-(e)] and Co layers (f) of sample A. The field of view
is about 25 $\mu$m and the spatial resolution 0.3 $\mu$m. The
projection of the x-ray incidence direction on the sample surface is
pointing up in the images (parallel to the arrow) and is parallel
(anti-parallel) to the direction of the field for positive
(negative) pulses. The magnetization direction is in the plane of
the layers and points up (parallel to the arrow) for black domains,
and down for white domains. The images were taken for delays between
photon and magnetic pulses of -15, 30, 55 and 80 ns, as indicated in
(a). The clearest visible black domains in the Co layer are
indicated with thin white lines in (f) and in the Fe image of
Fig.~(b) for comparison}
\end{figure}

For the time-resolved XMCD-PEEM imaging, we first induced a magnetic
domain structure in the Co layer before investigating the permalloy
magnetization reversal. This allows to study, in a same image
series, the behavior of the permalloy layer for reversal against and
with the local coupling direction, as well as the effect of the Co
domain walls. For the image series of sample A in Figure~2, the
magnetization of the sample was first saturated in the negative
direction. Then bipolar magnetic pulses were applied with the
temporal shape shown in Fig.~2~(a) and maximum and minimum field
strengths of 5.4~mT and -2~mT. The repeated application of these
pulses led after a short time to the zero field domain structures
shown in Fig.~2~(b) for permalloy and in Fig.~2~(f) for Co. After
these first few pulses, the Co domain structure remained stable
\cite{vogelMMM}. Images were then acquired for different delays and
representative images of the FeNi domain structure, taken for the
delays indicated in Fig.~2~(a), are shown in Fig.~2~(b)--(e). The Co
domain image in Fig.~2~(f) was recorded for a delay of -15~ns, but
the domain structure was the same for other delays. The magnetic
contrast in these time-resolved images was identical to the one of
static images, indicating reproducible reversal for this sample. The
reversal of the permalloy layer magnetization takes place mainly
through propagation of existing domain walls. The positive part of
the magnetic field pulses favors the growth of black domains
[Fig.~2~(c)], while negative fields favor the growth of white
domains [Fig.~2~(e)]. In Fig.~2~(d) the size of the white domains
increases compared to Fig.~2~(c), while the field is still in the
positive direction. In order to explain this the local coupling
between the permalloy and cobalt layers across the Cu spacer layer
has to be taken into account.

The largest and clearest black domains in the Co layer are
highlighted by thin white lines in Fig.~2~(f). These lines are
transposed to the permalloy image in Fig.~2~(b), at zero applied
field, for comparison. Clearly some correlation between the two
domain patterns exists, due to the magnetostatic coupling between
the layers. However, this coupling is not sufficient to induce a
parallel alignment between the local magnetization directions
everywhere in the two magnetic layers, and other parameters play a
role in determining the exact domain pattern of the permalloy layer.
In Fig.~2~(e), for example, some black domains persist in the
permalloy layer above white domains in the Co layer, even if in that
case both the external field and the local coupling with the Co
favor white domains. Some white domains are also pinned on large
defects that are clearly visible in the Co image. These results will
be discussed in more detail below.

\begin{figure}
\includegraphics*[bb= 173 352 424 518]{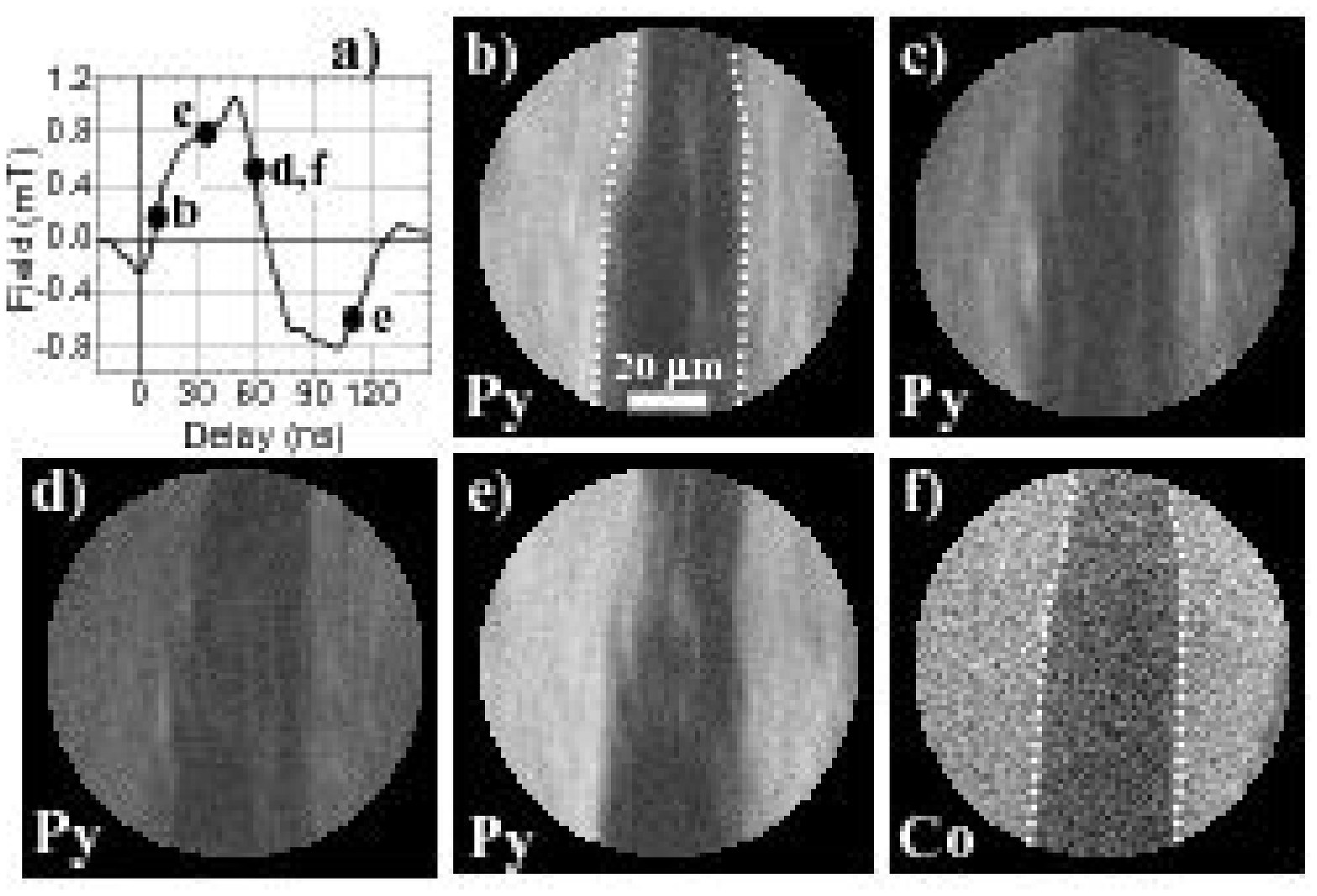}
\caption{Time and layer-resolved XMCD-PEEM domain images of the
permalloy [(b)-(e)] and Co (f) layers of sample B. The field of view
is about 80 $\mu$m and the spatial resolution 1 $\mu$m. The
direction of incoming photons and the magnetization directions of
the different domains are the same as in Fig.~2. The images were
taken at delays of 10, 35, 60 and 110 ns between photon and magnetic
pulses, as indicated in (a). The dotted white lines in (b) and (f)
indicate the position of the domain walls in the Co layer.}
\end{figure}

Time resolved XMCD-PEEM images for sample B are shown in Figure~3. A
domain pattern [Fig.~3~(f)] was first created in the Co layer using
1 ms long pulses from an external coil. Bipolar magnetic pulses with
the temporal shape shown in Fig.~3~(a) were then applied, with
maximum and minimum field strengths of about 1~mT and -0.8~mT. The
amplitude of these pulses was too small to affect the Co domain
pattern. At the beginning of the pulse, for an applied field close
to zero [Fig.~3~(b)], the correlation between the permalloy and Co
domain structures is much larger than in sample A. When the field
increases, regions in the permalloy that are white before the pulse
become almost homogeneously gray instead of showing a domain
structure [Fig.~3~(c) and 3~(d)]. This is caused by a
non-reproducibility of the permalloy domain structure, as
demonstrated by static images of the FeNi layer acquired after the
application of single magnetic pulses (Fig.~4). In this case, the Co
layer presented a black domain on the left and a white domain on the
right, separated by a domain wall as indicated on the permalloy
images with a thin white line. Images were obtained after
application of 1, 9 and 14 pulses (for Fig.~4~(a), (b) and (c),
respectively) with shape and amplitude as in Fig.~3~(a). The domains
are smaller than those obtained after application of quasi-static
pulses, but much larger than the domains in sample A. The domain
walls travel over relatively large distances (tens of microns) upon
reversal, and the domain structure is much less reproducible than in
sample A. In pump-probe mode, the obtained images are an average
over all different domain configurations, explaining the gray
contrast observed for the permalloy layer in Fig.~3.

\begin{figure}
\includegraphics*[bb= 180 432 414 509]{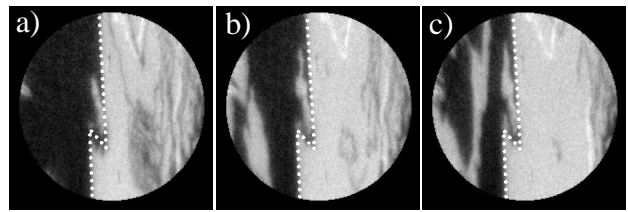}
\caption{Static XMCD-PEEM domain images of the permalloy layer of
sample B, taken after the application of 1, 9 and 14 pulses as shown
in Fig.~3~(a). Two domains are present in the Co layer, a black
domain on the left and a white domain on the right, separated by a
domain wall indicated in the images by a dotted white line.}
\end{figure}

A comparison between the different results shows that domain
patterns, the reproducibility and the correlation between permalloy
and Co domain structures during nanosecond switching of the
permalloy layer are very different for both samples. Image series
recorded with different pulsed field amplitudes and starting from
different domain patterns in the Co layer for both samples did not
show qualitative differences in the magnetization reversal of the
permalloy layer with respect to the representative series shown
here. Since both samples were deposited on the same substrates using
the same technique and their quasistatic coercivity and interlayer
coupling strength are quite similar, we attribute these differences
in their dynamic behavior to the difference in magnetic anisotropy.

In sample A the domains are some microns small and irregular. For
magnetic media with perpendicular anisotropy, the regularity of the
domain shape is determined by the domain wall stiffness
\cite{Thiele69,Shieh87,Kisi03}, which increases with increasing
anisotropy. In sample A, the weak magnetic anisotropy within the
plane leads to a relatively small domain wall stiffness. Domain
walls are easily distorted by pinning centers, leading to the
irregular domain shapes observed in Fig.~2 and to the formation of a
large number of 360${^{\circ}}$ domain walls \cite{vogelMMM}. The
application of short, strong magnetic field pulses increases this
tendency by increasing the nucleation density \cite{Pennec04}. As a
result, the magnetization is hard to saturate after application of
short pulses, even using quasi-static pulses. The local coupling
with the Co layer across the Cu spacer influences the local
magnetization reversal of the permalloy layer, but due to the large
deformability of the domain walls the detailed domain pattern is
determined mainly by local pinning. The geometry of the domains and
the distance between them also plays a role in the reversal, as we
showed in another paper \cite{KuchAPL}.

The domains in sample B are much larger than in sample A. The domain
walls are essentially straight and parallel to the easy axis. Domain
walls parallel to the magnetization direction accumulate less
magnetic charges and are therefore less energetic than domain walls
in other directions. The larger domain wall stiffness than in sample
A, due to the magnetic anisotropy, apparently causes pinning centers
to play a less important role in determining the domain shape during
magnetization reversal, which is dominated by magnetostatic effects
like the local coupling with the Co layer.

The domain pattern observed during reversal is much less
reproducible in sample B than in sample A. In general, domain walls
propagate by Barkhausen jumps, from a couple of pinning centers to
the next. These jumps can be quite reproducible when domain walls
move back and forth over small distances \cite{urbach95} like in
sample A. In sample B, the more rigid domain walls can be blocked by
pinning centers, until the accumulated Zeeman energy is large enough
to overcome the pinning barrier. By then, the domain wall will have
acquired enough energy to overcome a great number of pinning
centers, giving rise to Barkhausen 'avalanches' that are less
reproducible.

We can now also explain why much stronger pulses had to be used in
sample A (Fig.~2) than in sample B (Fig.~3), while the quasi-static
coercivities are very similar. In the hysteresis loops the permalloy
layer is initially saturated and domains have to nucleate before the
magnetization can reverse. When domains are already present, like in
the time-resolved PEEM measurements, the relevant magnetic field is
the one needed to cause domain wall propagation and this field is
smaller for sample B than for sample A. Moreover, coercivity and
saturation fields strongly increase upon increasing the applied
field sweep rate \cite{Gyorgy57,Pennec04}. We have performed Kerr
loop measurements for different field sweep rates indicating that
the increase of the saturation field going from quasi-static
measurements to nanosecond field pulses is much larger for sample A
than for sample B. We attribute this faster increase to the larger
domain wall pinning and the formation of many 360${^{\circ}}$ domain
walls.

In summary, we have used XMCD-PEEM to study magnetization reversal
in spin-valve like trilayers with spatial, temporal and layer
resolution. We show that magnetic anisotropy has an important
influence on the soft layer magnetization reversal. In the presence
of a uniaxial magnetic anisotropy within the plane the correlation
between the domain structures in the soft and hard magnetic layers
is strong. When this anisotropy is absent or very small, local
pinning effects mainly determine the domain structure upon reversal.

These results also illustrate why in applications using fast
magnetization switching the presence of magnetic anisotropy within
the plane of the layers should be preferred. For a soft magnetic
layer without such a magnetic anisotropy, the application of short
magnetic pulses induces the formation of small domains and of many
360${^{\circ}}$ domain walls. This leads to a large increase of the
field needed to magnetically saturate the sample.

We thank A.~Vaur\`{e}s and Y.~Conraux for sample preparation.
Financial support by BMBF (no.~05,~KS1EFA6), EU~(BESSY-EC-HPRI
Contract No.~HPRI-1999-CT-00028) and the Laboratoire Europ\'{e}en
Associ\'{e} 'Mesomag' is gratefully acknowledged. J.C. acknowledges
support through a "Ram\'{o}n y Cajal" contract and through project
No. MAT2003-08627-C02-02 from the Spanish Ministry of Science and
Technology.

\end{document}